\documentclass[twocolumn,superscriptaddress,showpacs,preprintnumbers,amsmath,amssymb]{revtex4}
\usepackage{graphicx}% Include figure files
\usepackage{dcolumn}% Align table columns on decimal point
\usepackage{bm}% bold math
\usepackage{color}
\usepackage{soul}
\usepackage{ulem}

\begin{document}

%\preprint {manuscript for Phys. Rev. B}

\date{\today}% It is always \today, today,

\title{Surface resonance of the Heusler half metal Co$_2$MnSi probed by SX-ARPES}
\author{Christian Lidig}
	\affiliation{Institut f\"ur Physik, Johannes Gutenberg-Universit\"at Mainz, 55099 Mainz, Germany}    
\author{Jan Min\'ar}
	\affiliation{New Technologies-Research Center, University of West Bohemia, Univerzitni 8, 306 14 Pilsen, Czech Republic}
\author{J\"urgen Braun}
	\affiliation{Department Chemie, Ludwig-Maximilians-Universit\"at M\"unchen, Butenandtstrasse 11, 81377 M\"unchen, Germany}
\author{Hubert Ebert}
	\affiliation{Department Chemie, Ludwig-Maximilians-Universit\"at M\"unchen, Butenandtstrasse 11, 81377 M\"unchen, Germany}
\author{Andrei Gloskovskii}
	\affiliation{Deutsches Elektronen-Synchrotron DESY, 22607 Hamburg, Germany}
\author{Jonas A. Krieger}
	\affiliation{Paul Scherrer Institut, CH-5232 Villigen PSI, Switzerland}
	\affiliation{Laboratorium für Festkörperphysik,  ETH Zürich, CH-8093 Zürich, Switzerland}
\author{Vladimir Strocov}
	\affiliation{Paul Scherrer Institut, CH-5232 Villigen PSI, Switzerland}     
\author{Mathias Kl\"aui}
	\affiliation{Institut f\"ur Physik, Johannes Gutenberg-Universit\"at Mainz, 55099 Mainz, Germany}
\author{Martin Jourdan}
	\affiliation{Institut f\"ur Physik, Johannes Gutenberg-Universit\"at Mainz, 55099 Mainz, Germany}   

\date{\today}
            
\begin{abstract}
Heusler compounds are promising materials for spintronics with adjustable electronic properties including 100\% spin polarization at the Fermi energy. We investigate the electronic states of AlO$_x$ capped epitaxial thin films of the ferromagnetic half metal Co$_2$MnSi ex-situ by soft X-ray angular resolved photoemission spectroscopy (SX-ARPES). Good agreement between the experimental SX-ARPES results and photoemission calculations including surface effects was obtained. In particular, we observed in line with our calculations a large photoemission intensity at the center of the Brillouin zone, which does not originate from bulk states, but from a surface resonance. This provides strong evidence for the validity of the previously proposed model based on this resonance, which was applied to explain the huge spin polarization of Co$_2$MnSi observed by angular-integrating UV-photoemission spectroscopy.  
\end{abstract}
\pacs{74.50.Cc, 79.60.-i}% PACS, the Physics and Astronomy
\keywords{Heusler, half metals, Photoemission}%Use showkeys class option if keyword
\maketitle

%%%%%%%%%%%%%%%%%%%%%%%%%%%%%%%%%%%%%%%%%%%%%%%%%%%%%%%%%%%%%%%%%%%%%%%%%%%%%%%%%%%%%%%%%%%%%%%%%%%%%%%
\section{INTRODUCTION}
The design and control of specific electronic properties of metallic thin films is a major requirement for the development of powerful spin based electronics (spintronics). Due to their compositional tunablity Heusler compounds represent a prime example for such an optimization of electronic states, which is usually based on band structure calculations. 

Angular resolved photoemission spectroscopy (ARPES) is a well established method providing in principle direct access to the electronic states. However, specifically in the case of Heusler compounds this proved to be very challenging as these materials do not cleave well, which is the standard method for the preparation of clean sample surfaces for ARPES. Additionally, their surfaces are in general highly reactive resulting easily in degraded surface regions of the samples.

The half metallic ferromagnet Co$_2$MnSi \cite{Gal02,Mei12} represents a typical example for this class of materials. Photoemission spectroscopy investigation of this and other Heusler compounds focused mainly on measurements of the spin polarization. Whereas the investigation of ex-situ prepared samples resulted in small polarization values \cite{Wan05, Fet13} only, in-situ spin-resolved and angular integrated UV-photoemission spectroscopy of epitaxial Co$_2$MnSi thin films demonstrated a large spin polarization close to 100\%\cite{Jou14}. This large spin polarization was measured in a range of binding energies much broader than expected by bulk band band structure calculations, which is consistent with photoemission calculations including surface effects and predicting a highly spin polarized surface resonance at the Fermi energy. However, direct experimental evidence for this surface resonance is required, as the angular integrated photoemission data did not show sufficient characteristic features for a detailed comparison with theory.

Without spin analysis the investigation of ex-situ prepared capped thin film samples has proven to be possible by less surface sensitive hard X-ray photoemission spectroscopy (HAXPES) \cite{Miy09,Oua11,Lid18}, but up to now no photoemission experiments on any Heusler compound, which show dispersive electronic states, are available.

Here we demonstrate that by ex-situ soft X-ray angular resolved photoemission spectroscopy (SX-ARPES) dispersive electronic states are observed investigating epitaxial Co$_2$MnSi(001) thin films capped by $1.8$~nm of AlO$_x$. The obtained experimental data allows for a more reliable test of the band structure calculations than the comparison with the previous angular integrated photoemission experiments \cite{Jou14, Bra15}. We now provide strong direct evidence for the validity of the calculated photoemission spectra previously used to explain the close to 100\% spin polarization observed in angular integrated UV-photoemission \cite{Jou14,Bra15}. We show that for Co$_2$MnSi also in the soft X-ray regime ($600-1200$~eV excitation energy) a surface resonance dominates the ARPES signal.

\section{TECHNIQUES}
Our epitaxial Co$_2$MnSi(001) Heusler thin films were prepared by rf-sputtering on MgO(001) substrates and capped by $1.8$~nm of AlO$_x$ as described elsewhere \cite{Lid18}.

SX-ARPES investigations of these samples were performed at the ADRESS beamline of the Swiss Light Source at the Paul Scherrer Institute \cite{Str10} using circularly polarized soft X-ray photons. The experimental geometry is associated with a grazing incidence of the X-rays \cite{Str10}. With a hemispherical energy analyzer (Specs PHOIBOS~150) the photoemission intensity was recorded as a function of the binding energy and momentum parallel to the thin film surface. The measurements were carried out at 12~K to reduce thermal effects diminishing the coherent k-resolved spectral component at high photoelectron energies \cite{Bra13}.

The experimental results were compared with photoemission calculations based on ab initio spin-density functional theory with local density
approximation. As in our previous work discussing angular integrated photoemission spectroscopy on Co$_2$MnSi, the electronic structure of
Co$_2$MnSi(001) was computed for a semi-infinite system including surface related effects using the fully relativistic Korringa-Kohn-Rostoker
method as implemented in the Munich SPR-KKR package \cite{Ebe17,Ebe11}. All technical details can be found in Refs.\,\cite{Jou14, Bra15}.

\section{SX-ARPES experiments and calculations}
For the identification of the center of the Brillouin zone ($\Gamma$-point) the photon energy was scanned in an energy range from
580 to 1200~eV. The corresponding plot of the photoemission intensity integrating from $E_b=-0.2$~eV to the Fermi energy is shown
in Fig.\,1 (right panel) in comparison with corresponding one-step model photoemission calculations (left panel) as described above.
\begin{figure}[htb]
\centering
		\includegraphics[width=0.80\columnwidth]{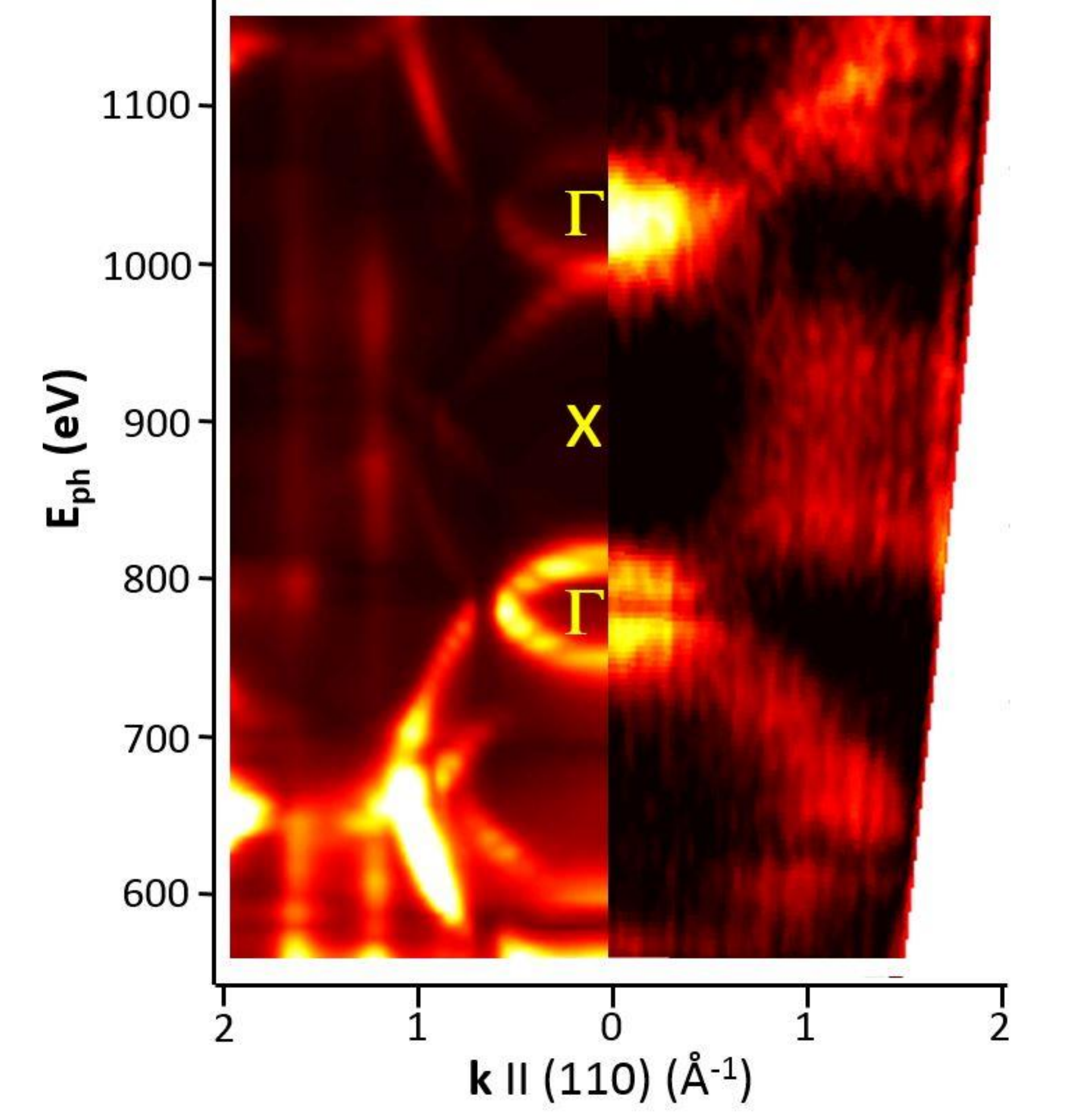}
	\caption{Photon energy and momentum $k\parallel (110)$ dependence of the photoemission intensity at the Fermi energy obtained investigating an epitaxial Co$_2$MnSi(001) thin film. The left part of the image represents the calculation, the right part the experimental photoemission data.}
\end{figure}
This corresponds to a cut through the Fermi surface perpendicular to the sample surface, as for a given binding energy the photon
energy selects the component $k_z$ of the crystal momentum perpendicular to the sample surface. Within a simple model, assuming a
free electron like final state of the photo excitation process, the photon energy can be converted into  $k_z$\cite{Huef, ADRESSE}.
However, within our one-step model photoemission calculations no such approximations are required and the photon energy dependence
of the ARPES intensity is calculated directly as shown in Fig.\,1.
Good agreement of the calculated (left panel) and experimental data (right panel) is found allowing the identification of the center
of the Brillouin zone ($\Gamma$-point), which corresponds to a photon energy of 1020~eV. 

To add further evidence for the validity of our combined band structure and photoemission calculations, we additionally show the emission
angle dependence of the SX-ARPES intensity obtained with a photon energy of 1020~eV and integrating from $E_b=-0.2$~eV to the Fermi energy.
This  corresponds to a cut through the Fermi surface in a reciprocal space plane parallel to the sample surface through the $\Gamma$ point.
In Fig.\,2, the experimental data (bottom panel) is shown next to the corresponding calculated intensity distribution (top panel). Again,
good agreement concerning the large spectral weight at the $\Gamma$-point is obtained.
   
\begin{figure}[htb]
\centering
		\includegraphics[width=0.65\columnwidth]{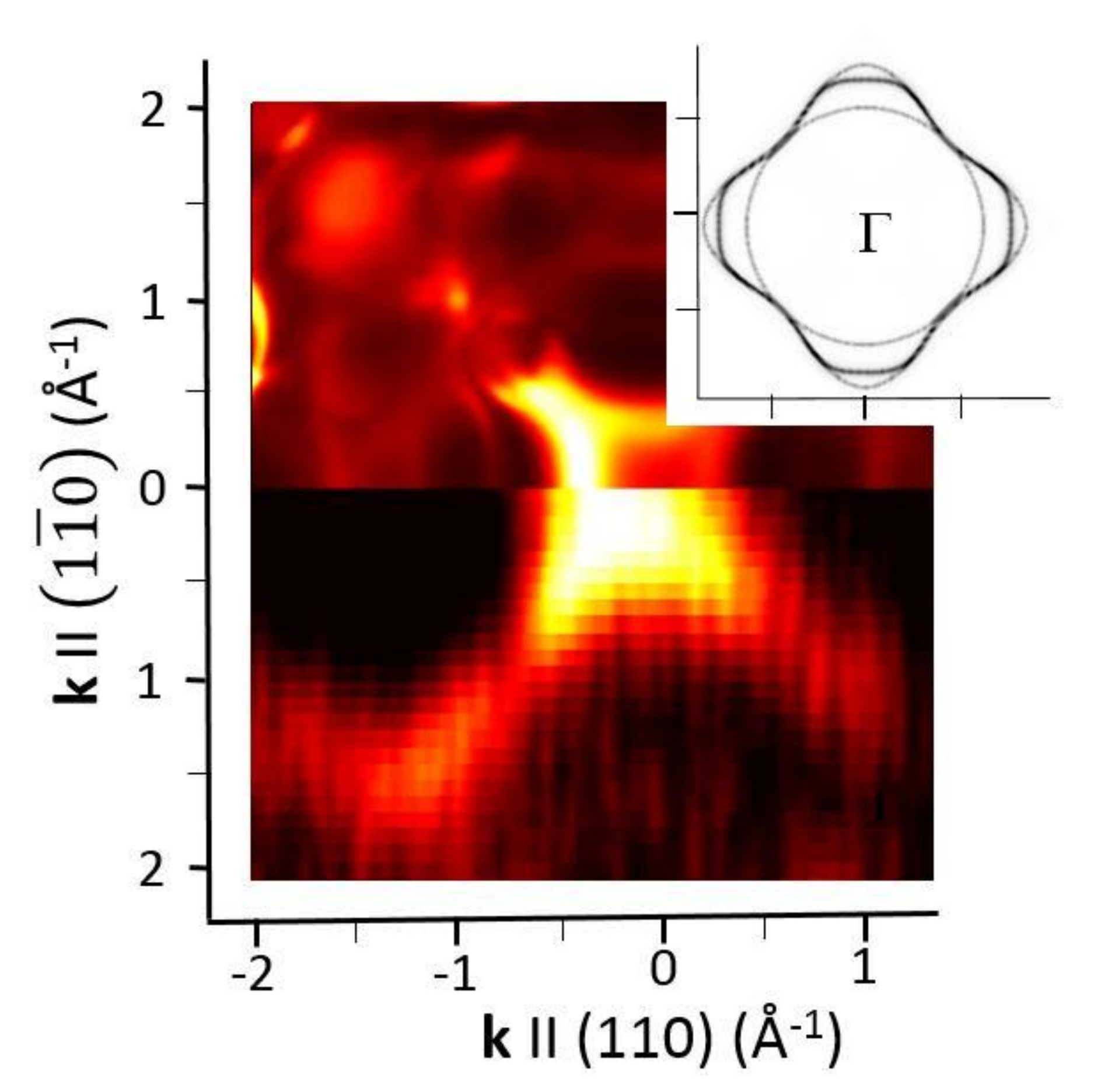}
	\caption{Cut through the Fermi surface of Co$_2$MnSi(001) in a reciprocal space plane parallel to the sample surface through the $\Gamma$-point. Top panel: calculation including surface effects; bottom panel: experimental photoemission data. The inset (same scale as main figure) shows the corresponding calculated bulk Fermi surface of Co$_2$MnSi.}
\end{figure}

Finally, in Fig.\,3 we show dispersing electronic valence states with momentum $k\parallel (110)$, i.\,e.\,along the $\Gamma$-K direction (upper panel) and for $k\parallel (100)$, i.\,e.\,along the $\Gamma$-X direction (lower panel). Specifically, close to the Fermi energy for binding energies below $-1$~eV good agreement between experimental data (right) and calculation (left) is obtained, which is consistent with the good agreement of the calculated and measured Fermi surface cuts.

For larger binding energies $>1$~eV no comparison of the experimental data with the calculations is possible, as no dispersive features are observed by SX-ARPES. Instead, at a binding energy of $1$~eV a k-independent large photoemission intensity is experimentally observed, which could be associated to photoemission from the AlO$_x$ capping layer or to magnon scattering.

\section{Surface resonance vs.\,bulk states}

At the Fermi energy we consistently observe in agreement with our photoemission calculations a large photoemission intensity at the $\Gamma$-point, which vanishes with increasing values of $k$ for both directions parallel (110) and (100) to energies above the Fermi edge. However, in most published calculations of the band structure of Co$_2$MnSi no electronic bulk states are present at $E_F$ in the region around the $\Gamma$-point. Typically, the calculated bulk majority bands cross the Fermi energy along the $\Gamma$-X direction, i.\,e.\,for $k \parallel (100)$ close to the X-point, whereas the minority states show a gap with the Fermi energy either in the center or close to the edge of the band gap \cite{Ish98,Ko10,Bai13,Fet15,Lan16}. This is consistent with our own calculations of the bulk band structure of Co$_2$MnSi and we obtain a bulk Fermi-surface with a radius of $\simeq 0.7$~$\rm{\AA}^{-1}$ as shown in the inset of Fig.\,2.

\begin{figure}[htb]
\centering
		\includegraphics[width=0.9\columnwidth]{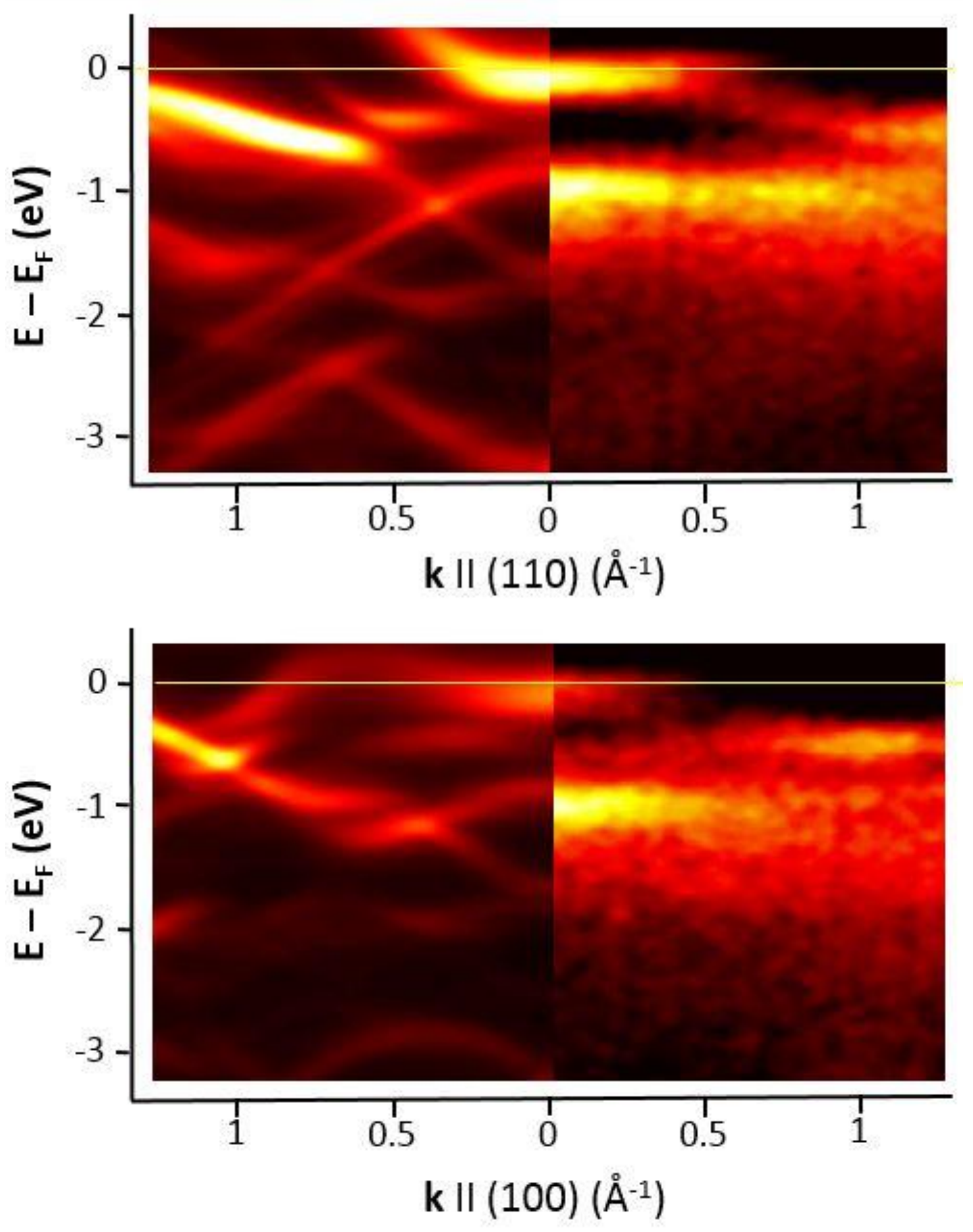}
	\caption{Dependence of the calculated (left panel) and experimental (right panel) photoemission intensity on the binding energy $E-E_F$ and on the photoelectron momentum parallel to the sample surface. Top panel: $k\parallel$(110); bottom panel: $k\parallel$(100).}
\end{figure}

Nevertheless, as the minority states at the $\Gamma$-point in the bulk band structure are close in energy to $E_F$, they in principle might show up in the experimental SX-ARPES Fermi surface due to energy broadening effects. Also the surface resonance discussed in our previous work to explain the close to 100\% spin polarization observed by spin-resolved angle-integrating UV-photoemission spectroscopy (SRUPS) \cite{Jou14} is situated close to, but above, the Fermi energy of Co$_2$MnSi \cite{Bra15}. In this framework it is important to note, that the exact position of the Fermi energy is in general difficult to obtain by band structure calculations. Here, the comparison with the experimental data shown in Fig.\,3 serves to determine the Fermi energy in our band structure calculations. We deduce, that the surface resonance is situated about $0.15$~eV above the Fermi energy, whereas the bulk minority states are found about $0.25$~eV below the Fermi energy. Please note that this corresponds to a shift of the Fermi energy by about $0.25$~eV to higher energies compared to our previous estimation based on the SRUPS results with relatively poor energy resolution \cite{Bra15}.      

Additional to being closer in energy to E$_F$ as the bulk minority states, the surface resonance also contributes much stronger to the SX-ARPES intensity than all other electronic states. To elucidate it's character we calculated the layer-dependence of the SX-ARPES intensity: A prototypical surface resonance splits up in energy from the corresponding bulk bands and disperses in the vicinity of these bulk states. The resonance shows up with a considerable spectral weight within the first three or four atomic layers. Due to its bulk contribution \cite{Bra14}, the spectral weight of a resonance is in general much larger than that of a real surface state. Thus, the surface resonance can be observed even with soft X-ray excitation of about 1~keV. At these energies the information depth of about $\simeq 25$~$\rm{\AA}$ \cite{Tan11} allows for a layer-dependent study of the spectral distribution of the various electronic states in order to estimate the fraction surface and bulk-like contributions to the total intensity distributions. Correspondingly, we performed layer-resolved one-step photoemission calculations in normal emission to reveal the inelastic mean free path (IMFP) of the surface resonance.

The result is shown in Fig.\,4, where the absolute value of the photoemission intensity is plotted as a function of the atomic layer number starting from the surface (layer 0). This is shown for different initial-state energies, from the energy of the surface resonance $E_{SR}$ to larger binding energies. It is clearly visible, that the spectral weight in the first four layers is considerably enhanced. For initial-state energies close to $E_{SR}$ the surface resonance shows up with the maximum intensity in the photoemission spectrum. For other energies the IMFP is estimated to about 10-12 layer in accordance with \cite{Tan11} and as a consequence indicates bulk-like emission. The absolute value of the spectral weight is peaked at layer 2, with a considerably contribution at the fourth layer, where even layer numbers indicate layers containing Co atoms only. This way the surface resonance could be identified as a Tamm-like surface feature that split up in energy from the corresponding Co bulk states.
\begin{figure}[htb]
\centering
		\includegraphics[width=0.9\columnwidth]{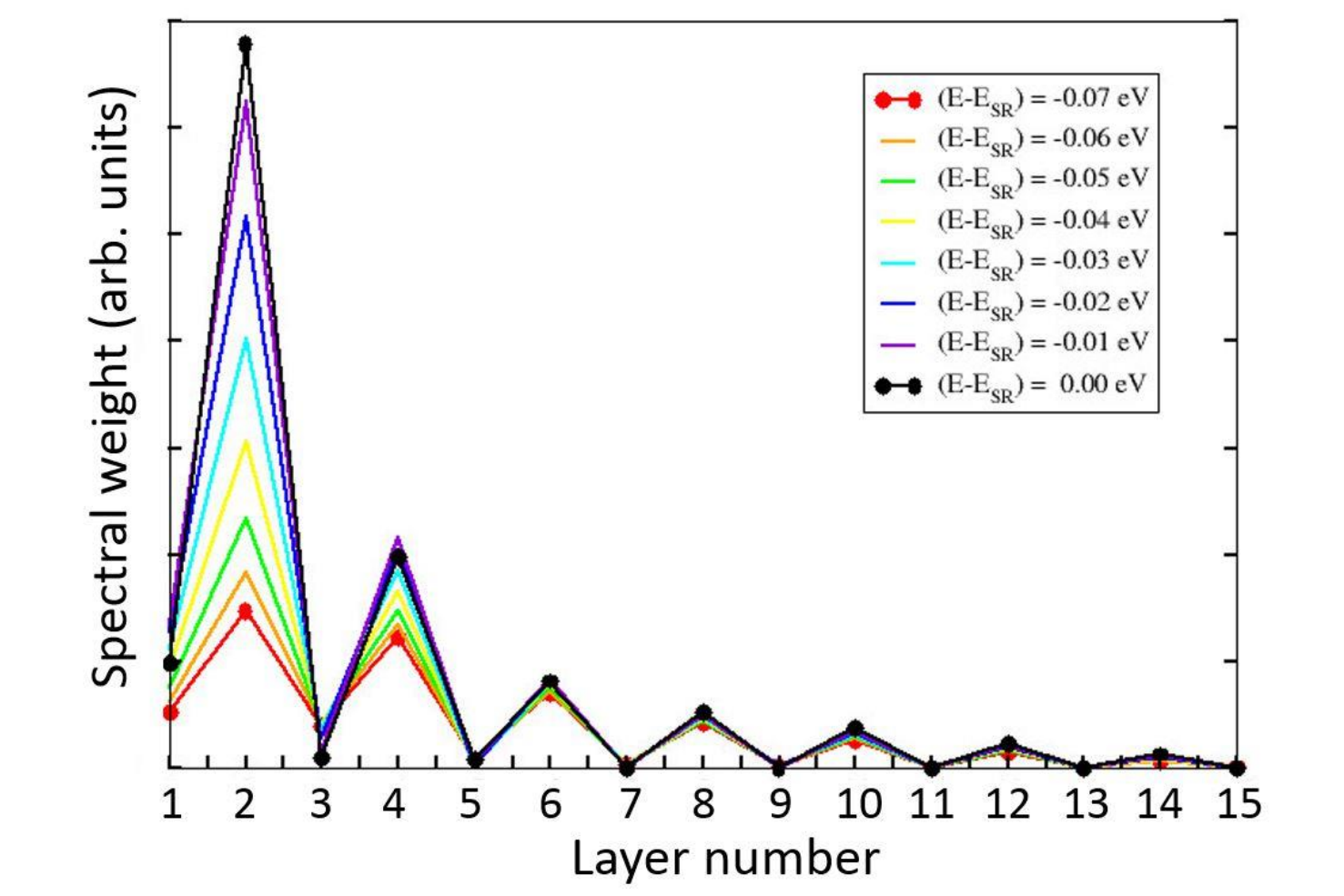}
	\caption{Spectral weight of different electronic states plotted as a function of the initial-state energy, from the energy $E-E_{SR}=0$~eV (black), indicating the maximum in the amplitude of the surface resonance intensity distribution, to $E-E_{SR}=-0.07$~eV (red), corresponding to bulk-like emission.}
\end{figure}

\section{Summary}
In summary, the Fermi surface of the half metallic Heusler compound Co$_2$MnSi as well as dispersive bands close to the Fermi energy were investigated by SX-ARPES of epitaxial thin films capped by a $1.8$~nm AlO$_x$ layer. The experiments are compared with photoemission calculations including surface related effects, which results in good agreement. In particular, a large photoemission intensity is obtained at the center of the Brillouin zone, although no electronic state are present around the $\Gamma$-point in bulk band structure calculations. Based on our photoemission calculations of the semi-infinite system this is explained by a surface resonance of Co$_2$MnSi(001) dominating the SX-ARPES intensity. The comparison of theory and experiments based on angular resolved data provides direct strong  evidence for the existence of this resonance proposed previously to explain the close to 100\% spin polarization observed by angular integrated photoemission experiments \cite{Jou14, Bra15}. 

\textit{Financial  support  by  the  the  German  Research  Foundation (DFG) via projects Jo404/9-1 and Eb158/32-1 is acknowledged. JAK was supported by the Swiss National Science Foundation (SNF-Grant No. 200021 165910).}

\end{document}